\documentclass[preprint]{iucr}              

     \journalcode{J}              

\usepackage{siunitx}
\usepackage{geometry,color,graphicx}

\begin{document}                  



\title{Pulse-to-pulse wavefront sensing at free-electron lasers using ptychography}


\author[a,b]{Simone}{Sala}\aufn{Current address: MAX IV Laboratory, Lund University, Lund, \country{Sweden}}

\author[c,d]{Benedikt J.}{Daurer}
\author[e]{Michal}{Odstrcil}
\author[f]{Flavio}{Capotondi}
\author[f]{Emanuele}{Pedersoli}
\author[g]{Max F.}{Hantke}
\author[f]{Michele}{Manfredda}
\author[d,g]{N. Duane}{Loh}

\cauthor[b]{Pierre}{Thibault}{pierre.thibault@soton.ac.uk}

\cauthor[c]{Filipe R. N. C.}{Maia}{filipe@xray.bmc.uu.se}

\aff[a]{Department of Physics \& Astronomy, University College London, London, \country{UK}}
\aff[b]{Department of Physics \& Astronomy, University of Southampton, Southampton, \country{UK}}
\aff[c]{Department of Cell and Molecular Biology, Uppsala University, Uppsala, \country{Sweden}}
\aff[d]{Department of Biological Sciences, National University of Singapore, \country{Singapore}}
\aff[e]{Paul Scherrer Institut, Villigen, \country{Switzerland}}
\aff[f]{Elettra-Sincrotrone Trieste, Trieste, \country{Italy}}
\aff[g]{Department of Chemistry, Oxford University, Oxford, \country{UK}}
\aff[h]{Department of Physics, National University of Singapore, \country{Singapore}}



\maketitle                        


\begin{abstract}
The pressing need for the detailed wavefront properties of ultra-bright and ultra-short pulses produced by free-electron lasers (FELs) has spurred the development of several complementary characterization approaches. Here we present a method based on ptychography that can retrieve full high-resolution complex-valued wave functions of individual pulses. Our technique is demonstrated within experimental conditions suited for diffraction experiments in their native imaging state. This lensless technique, applicable to many other short-pulse instruments, can achieve diffraction-limited resolution.
\end{abstract}


\section{Introduction}

Free-electron lasers (FELs) are opening the way to a number of new research paths. Within the field of microscopy, the highly coherent and short pulses produced by FELs are used to conduct diffractive imaging of individual particles, also called flash X-ray imaging (FXI), potentially down to atomic resolution \cite{Neutze2000, Chapman2006a, Seibert2011}. Many other investigations exploit FEL tight focal spots to maximize fluence or improve spatial resolution \cite{Willems2017, Vidal2017, Mincigrucci2018}. For all these applications, a reliable high-resolution characterization of the shot-to-shot focal spot is crucial. A number of beam diagnostics methods have been purposely designed for this task. However, with ultrashort pulses (down to femtoseconds) which can reach a flux sufficient to destroy or irreversibly damage most targets, these methods mostly provide only partial information about the wavefront, such as position, size, shape, intensity or curvature \cite{Chalupsky2011, Vartanyants2011, Rutishauser2012, Loh2013, Sikorski2015, Keitel2016, Daurer2017}. A recent grating-based method can provide real-time wavefront distributions \cite{Schneider2018, Liu2018}, though away from the focal plane and with a resolution limited by the grating's manufacturing process.

At third-generation synchrotron sources, ptychography is now a popular wavefront characterization tool \cite{Kewish2010a, Schropp2010, Takahashi2011, Honig2011, Vila-Comamala2011}, thanks to its ability to retrieve the complex-valued wavefield in or close to the focal plane (the probe) along with the transmission function of the sample (the object)  \cite{Thibault2009, Maiden2009}. Recent improvements to ptychographic reconstruction algorithms address additional sources of data degradation, such as partial coherence \cite{Thibault2013}, scanning position jitter \cite{Guizar-Sicairos2008, Maiden2012a, Beckers2013, Zhang2013, Tripathi2014} and probe variations \cite{Odstrcil2016}. The recovered illumination wavefields can be numerically propagated to refine the focal position or to reveal optics-induced aberrations.

Here we show that ptychography can be used to reliably reconstruct the wavefront of individual FEL pulses. Our approach requires no \textit{a priori} information on either object or probes, and all individual wavefronts are allowed to vary, including their positions. Unlike a previous application of ptychography at the FEL which filtered the pulses through additional optics \cite{Schropp2013}, our method retrieves wavefronts in the native imaging state, i.e.\ the same conditions in which the apparatus is used for imaging or diffraction experiments.


\section{Methods}

\begin{figure}
\includegraphics[width=0.6\textwidth]{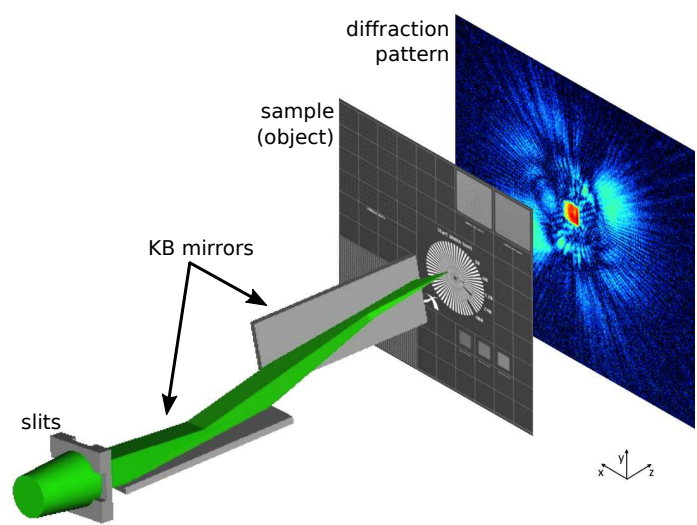}
\caption{Diagram of the experimental setup. Adjustable slits form a pupil that admits the central part of the FEL beam. Kirkpatrick-Baez mirrors focus the beam onto a small area of the sample, which is scanned with a translation stage in the $x$-$y$ plane. The intensity of the resulting free-space propagated exit waves (i.e.\ the diffraction patterns) are recorded by a detector downstream along $z$.}
\label{fig:setup}
\end{figure}

\begin{figure}
\includegraphics[width=0.7\textwidth]{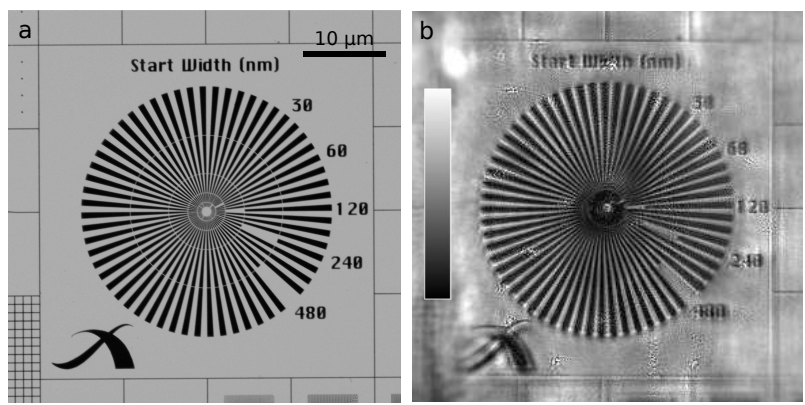}
\caption{SEM image (a) and amplitude of ptychographic reconstruction (b) of a Siemens star test pattern; the color bar in (b) represents transmission between 0 and 1.}
\label{fig:SEM vs ptycho}
\end{figure}

The experiment was carried out at the DiProI instrument of FERMI (see Appendix for details), a seeded FEL producing $\SI{10}{\micro\joule}$ pulses, at a photon energy of $\SI{83}{\electronvolt}$, equivalent to a wavelength of $\SI{15}{\nano\meter}$. Figure \ref{fig:setup} gives a schematic representation of the experimental setup. After attenuation by about 4 orders of magnitude to avoid damage, the beam was focused by a pair of perpendicular bendable Kirkpatrick-Baez (KB) mirrors to a focal spot size of under $10\times\SI{10}{\micro\meter}^2$. The sample -- positioned close the focal plane -- was a gold test pattern featuring a $30\times\SI{30}{\micro\meter}^2$ Siemens star whose SEM image is represented in Fig.\ \ref{fig:SEM vs ptycho}a. At $\SI{83}{\electronvolt}$ the sample behaves as a binary object since the gold-plated areas absorb completely the incident X-rays. Far-field diffraction patterns were detected with a CCD camera located $\SI{150}{\milli\meter}$ downstream from the sample.

Ptychographic data were collected by scanning the sample over the Siemens star area. A wide jitter of the photon beam position, observed throughout data acquisition and caused by vibrations of an upstream optical component, was first corrected using a cross-correlation approach and then further refined. This procedure, which would normally not be needed in a routine application of the method, reduced the number of valid frames down from 1515 to 937.

The ptychographic reconstruction was carried out using a single-pulse retrieval algorithm derived from the orthogonal probe relaxation ptychographic (OPRP) method \cite{Odstrcil2016} which recovers a different probe for each diffraction pattern while keeping the problem over-constrained through dimensionality reduction. Individual probes are thus modeled as linear combinations of a number of dominant components (also called ``modes'' or ``eigenprobes'') recovered dynamically and without {\it a priori} information.
As usual for ptychography, the object's transmission function is also retrieved without enforcing any prior knowledge. The essence of OPRP is compatible with any ptychographic reconstruction algorithm and has been implemented within the \textit{PtyPy} reconstruction suite \cite{Enders2016} for both difference map (DM) \cite{Thibault2009} and maximum likelihood (ML) \cite{Thibault2012} algorithms. The reconstructions presented here were obtained using DM followed by ML refinement, both using a 10-component decomposition of the retrieved probes.


\section{Results}

\begin{figure}
\includegraphics[width=0.97\textwidth]{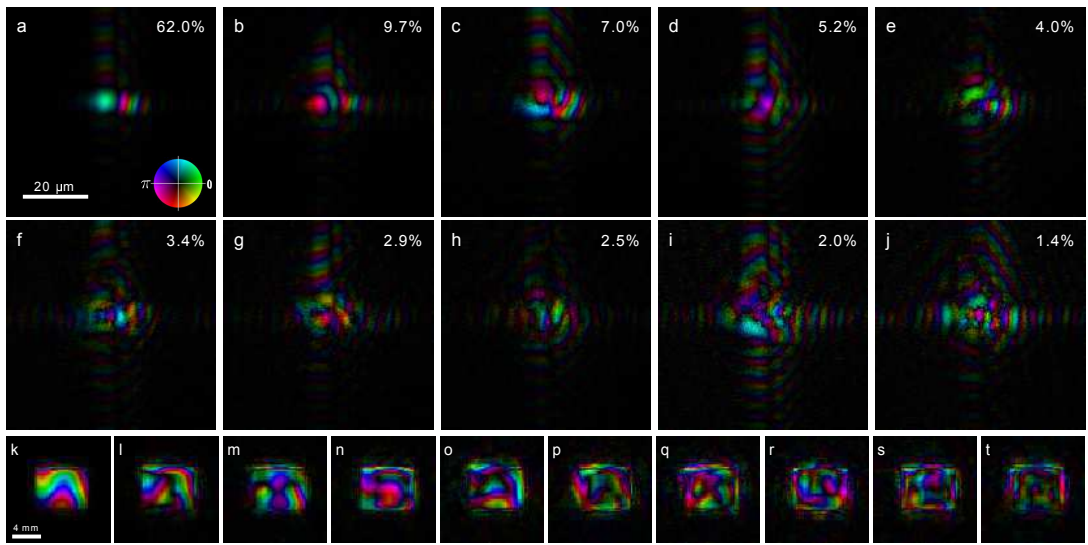}
\caption{(a-j) 10 components of reconstructed probes obtained via OPRP reconstruction. Normalised singular values are annotated on each component. (k-t) Back-propagation of same components to the pupil plane, neglecting spherical wave term. Amplitude is mapped to brightness and phase to hue according to the color wheel in (a); intensity scale is relative.}
\label{fig:modes}
\end{figure}

The absolute value of the retrieved object transmission function is represented in Fig.\ \ref{fig:SEM vs ptycho}b, which can be compared with the SEM image of the same region in Fig.\ \ref{fig:SEM vs ptycho}a. The agreement between the two images is apparent, even well outside the field-of-view of the original scanning area, thanks to the wide beam jitter effectively enlarging the imaged area. The fidelity of the reconstructed object confirms the robustness of the algorithm and the validity of the retrieved probes, whose components are represented in Fig.s \ref{fig:modes}a-j. Though each component's contribution to each pulse varies, a qualitative indication of their relative weight is given by the singular values obtained through truncated singular-value decomposition (SVD) and annotated on each component (cf Fig.s \ref{fig:modes}a-j). When back-propagated to the virtual secondary source plane located at the mid-point between the pair of KB mirrors, $\SI{1.48}{\meter}$ upstream from the interaction plane (Fig.s \ref{fig:modes}k-t) the components exhibit the expected intensity distribution, with the intensity dropping to negligible values outside of the main pupil.

\begin{figure}
\includegraphics[width=0.97\textwidth]{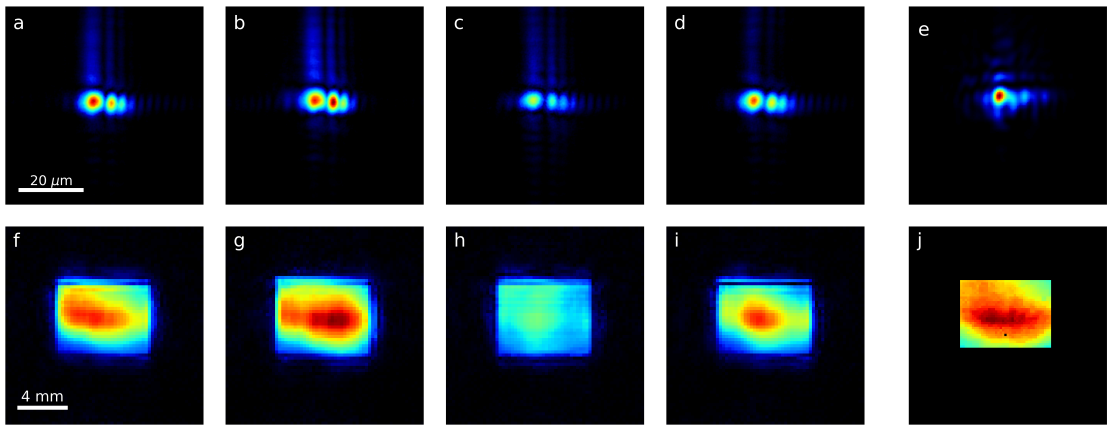}
\caption{(a-d) Normalized amplitude of 4 of the $N=937$ retrieved probes obtained via OPRP reconstruction. (e) Amplitude of probe retrieved from Hartmann sensor data. (f-j) Back-propagation of (a-e) to the virtual pupil plane, neglecting spherical wave term.}
\label{fig:probes}
\end{figure}

The normalized amplitude of four selected probes is shown in Fig.s \ref{fig:probes}a-d as a representative sample of the full stack of $N=937$ retrieved probes which is available as a video in the Supplementary Material. For comparison purposes, the amplitude of a probe retrieved starting from Hartmann sensor data is shown in Fig.\ \ref{fig:probes}e. The Hartmann sensor routinely available at the beamline for wavefront sensing \cite{Raimondi2013} was operated within the same experimental conditions as those of the ptychography experiment, although it did not sample the same pulses. Figures \ref{fig:probes}f-j show the amplitude of the same wavefronts (Fig.s \ref{fig:probes}a-e) after they have been numerically back-propagated to the virtual secondary source plane. Pulse-to-pulse variations can be observed both at the sample plane and at the virtual secondary source plane.

\begin{figure}
\includegraphics[width=0.6\textwidth]{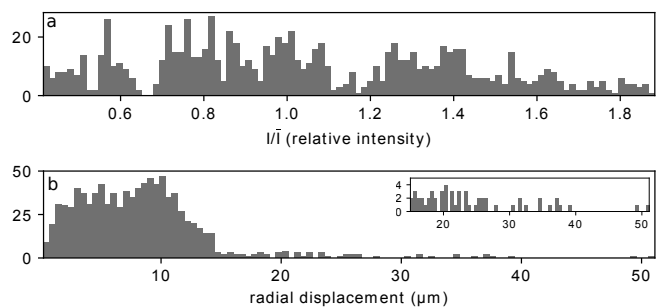}
\caption{Histograms of (a) intensity of each probe relative to the median intensity) and (b) radial displacement of the center of mass of each probe relative to center of the detector. Inset in (b) is a rescaled version of a portion of the same histogram.}
\label{fig:stat}
\end{figure}

Using the full stack of retrieved probes, it is possible to gather valuable statistical information, such as variations in intensity and beam pointing. Figure \ref{fig:stat}a illustrates the fluctuations of the total intensity found for each probe relative to the median total intensity, revealing significant variations, with a relative standard deviation of $0.4$. Figure \ref{fig:stat}b shows the radial displacement of the center of mass of each probe relative to the center of the detector, revealing a median relative displacement of $\SI{8}{\micro\meter}$, confirming the presence of beam pointing variations.

\begin{figure}
\includegraphics[width=0.95\textwidth]{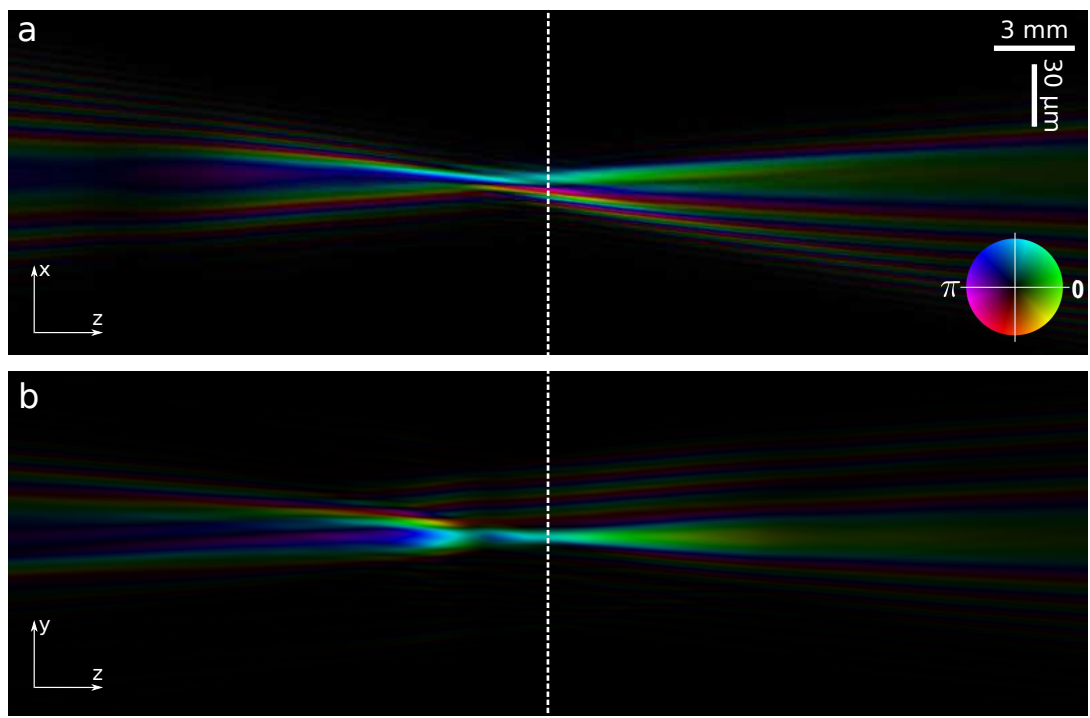}
\caption{Horizontal (a) and vertical (b) sections of ptychographic reconstruction of main component propagated around the focal position ($\pm$ 20 mm). Image scaling is different in its two dimensions according to the scale bars in (a). Amplitude is mapped to brightness and phase to hue according to the color wheel in (a).}
\label{fig:propagation}
\end{figure}

We numerically propagated the main component -- as it is the most representative part of all retrieved probes -- downstream and upstream from object plane to investigate the region around the focal position. Sections of the propagated main component are shown in Fig.\ \ref{fig:propagation} and can be compared with the unpropagated reconstructed main component from Fig.\ \ref{fig:modes}a. The dotted line in Fig.\ \ref{fig:propagation} indicates the sample plane; the focal plane has been estimated to be some $\SI{2}{\milli\meter}$ further upstream, as the plane at which the beam size is minimized.


\section{Conclusions}

We have demonstrated the first fully pulse-to-pulse ptychographic wavefront reconstruction for a FEL instrument in its native imaging state. Our approach provides the complex-valued wavefront of each pulse directly at the sample plane, in its near-focus position. As a lensless microscopy technique, its achievable resolution is limited only by flux and detector numerical aperture.

The full set of 937 retrieved probes reveals pulse-to-pulse variations, providing statistical information on FEL beam fluctuations and direct insight into FERMI's performance. The retrieved probes also confirm the elongated shape of the beam at and around the focal position, as expected from previous experiments carried out at the same beamline \cite{Raimondi2013}.

Ptychographic wavefront characterisation brings some benefits compared to grating-based methods \cite{Schneider2018, Liu2018}, which recover information far from the focal position, tend to underestimate the intensity in the tails of the power distribution and are limited by grating fabrication. The method has also proven effective even in the presence of significant setup vibrations -- another advantage over Hartmann and grating-based wavefront sensing, which are mostly insensitive to beam position variations.

Pulse-to-pulse ptychography is primarily expected to benefit wavefront characterization experiments at FELs, contributing to the development of FEL and FEL-based science. Its application can be extended to other imaging experiments, which would benefit from the relaxation of the single-illumination probe constraint.



\appendix

\section{Experimental setup}

The ptychography experiment was carried out at the Diffraction and Projection Imaging (DiProI) beamline \cite{Capotondi2013, Capotondi2015} at FERMI, EUV and soft X-ray seeded FEL, using the FEL-2 line \cite{Allaria2012, Allaria2013, Allaria2015}. The vertical and horizontal bendable KB mirrors had a focal length of $\SI{1.75}{\meter}$ and $\SI{1.2}{\meter}$, respectively \cite{Raimondi2013}. The sample (Xradia X30-30-2) was a $\SI{110}{\nano\meter}$ S$\text i_3\text N_4$ membrane with a $\SI{200}{\nano\meter}$ thick Au test pattern deposited on top. A 3-axis translation stage was used to translate and scan the sample within a vacuum chamber. The in-vacuum CCD camera was a PI-MTE:2048B with $2048\times2048$ pixels, $\SI{13.5}{\micro\meter}$ each. To decrease readout time to $\SI{2}{\second}$, only the intensities collected by the central $1000\times1000$ pixels were recorded. As readout frequency was lower than the FEL's $\SI{10}{\hertz}$ repetition rate, a fast shutter was used to prevent more than one pulse from contributing to each detector reading. Due to further overhead, the acquisition rate was effectively reduced to $\SI{0.2}{\hertz}$, i.e.\ only one every 50 FEL pulses was recorded.

The beam was attenuated with a combination of attenuators. A $\SI{6}{\meter}$ long gas chamber was filled with $2.8\times10^{-2}\,\SI{}{\milli\bar}$ of $\text N_2$ and complemented with solid Zr ($\SI{600}{\nano\meter}$) and Al ($\SI{200}{\nano\meter}$) attenuators. Thus attenuated, pulses remained well below the sample's damage threshold, given by Au melting dose of $\SI{0.4}{\electronvolt}$ per atom \cite{David2011}.

\section{Data acquisition and processing}

\begin{figure}
\includegraphics[width=0.85\textwidth]{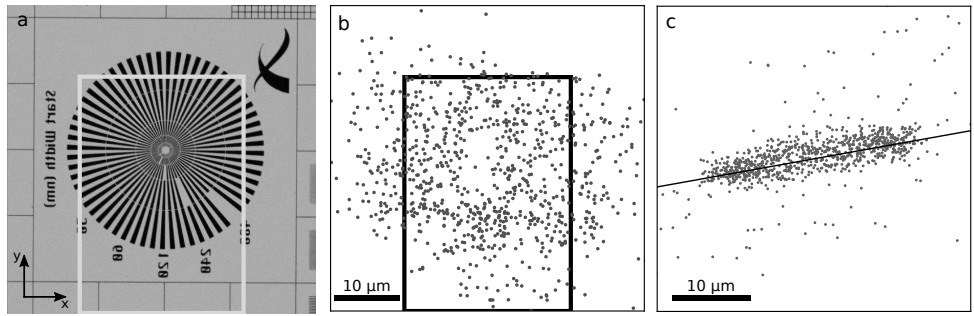}
\caption{(a) SEM image of the Siemens star test pattern; scalebar in (b). (b) Probe positions recovered via our position correction algorithm. The area covered by the motor positions used for the ptychographic scan is annotated with rectangles in (a) and (b). (c) Position correction for each scanning point; null correction (0,0) at center. The main axis of such two-dimensional correction distribution is annotated revealing a dominant vibration component along the $x$ axis.}
\label{fig:pos correction}
\end{figure}

Ptychographic data were collected using three $25\times\SI{25}{\micro\meter}^2$ spiral scans in the $x$-$y$ plane, with a step size of $\SI{2.5}{\micro\meter}$ for a total of 101 positions each. Five single-pulse diffraction patterns were collected at each scanning position, for a total of 1515 frames over the 3 spiral scans. The center of the last spiral scan was translated in $y$ to extend the overall scanned area which is highlighted by the rectangles in Fig.s \ref{fig:pos correction}a,b. The collected diffraction patterns were binned by a factor of 2 and then padded to a size of $512\times512$ pixels in order to decrease computational cost. Each frame was then dark-subtracted. Detector counts were thresholded to a value of 0 in order to remove unphysical negative counts and converted into units of photon counts.

A wide jitter of the photon beam position was found to have occurred throughout data acquisition. Ascribed to vibrations in the optics upstream from the experimental setup, the jitter was of such a large amplitude -- tens of microns -- that the scanning positions recorded by the sample motors were essentially unusable. Most known position refinement approaches within ptychographic algorithms \cite{Guizar-Sicairos2008, Maiden2012a, Beckers2013, Zhang2013, Tripathi2014} have been designed to account for minor deviations from the expected positions, typically smaller than the scanning step size. Here, a customized approach was implemented to accommodate for such a wide jitter. Coarse positions were first obtained by cross-correlating the recorded diffraction patterns with simulations based on the interaction of a simulated probe and an object modelled from the available high-resolution image of the test pattern (cf Fig.\ \ref{fig:pos correction}a). The coarse positions were then further refined with a purposely-designed position refinement algorithm \cite{Odstrcil2018}. Diffraction patterns corresponding to positions far from the intended field-of-view were discarded, as well as others for which position refinement failed, effectively reducing the dataset to 937 frames.

Figure \ref{fig:pos correction}b shows the corrected and refined positions, along with the intended rectangular scanning area. The deviation of the retrieved scanning positions to the nominal ones is represented in Fig.\ \ref{fig:pos correction}c where the main axis of the vibration distribution is annotated, revealing a dominant horizontal component associated to a standard deviation of $\SI{9.2}{\micro\meter}$.

Although the coarse position correction was only possible thanks to the in-depth characterization of the test pattern prior the experiment, the developed position refinement algorithm is expected to benefit several other experiments performed at FELs which are affected by intrinsic pointing instability, beside minor vibrations of the optics and sample stages.

Given the experiment geometry and detector specifications, the achieved pixel size in the ptychographic reconstruction was $\SI{162}{\nano\meter}$. The initial illumination function was produced by numerical back-propagation of the mean diffraction pattern. The initial object was assigned a uniform unit transmission function. The ptychographic algorithm ran 200 iterations of DM, followed by 800 of ML refinement.

\section{Reconstruction algorithm}

The ptychographic wavefront characterization approach used within this work for single-pulse investigation is derived from OPRP \cite{Odstrcil2016}. It recovers a different probe $P_j$ for each of the $N$ diffraction patterns $I_j$ recorded during a ptychographic scan, with the frame index $j$ varying between $1$ and $N$. The SVD step is added at the end of every iteration of the ptychographic reconstruction algorithm and generates the probes' main principal components or ``modes''.

Given the complex matrix $P$ whose columns contain estimates of the individual probes $P_j$, applying SVD to $P$ leads to $P=U\Sigma V^*$ where $V^*$ denotes the Hermitian transposition of $V$ and both $U$ and $V$ are unitary matrices such that $UU^*=U^*U=I$ and $V^*V=VV^*=I$ with $I$ as the identity matrix. By multiplying $P$ with its Hermitian transpose $P^*$, one gets
\begin{equation}
P^*P = V\Sigma^*U^*U\Sigma V^* = V(\Sigma^*\Sigma)V^*
\end{equation}
with $P^*P$ as a Hermitian matrix and $(\Sigma^*\Sigma)$ as a diagonal matrix. This is equivalent to the eigenvalue problem
\begin{equation}
(P^*P)V=V(\Sigma^*\Sigma)
\end{equation}
so that the non-zero elements on the diagonal of $\Sigma$ correspond to the square roots of the eigenvalues of $P^*P$.

The solution to this problem within the ptychographic algorithm is implemented as a truncated diagonalisation: the $N$ eigenvalues and main $k$ eigenvectors $\hat V$ can be retrieved, with $k<N$ and $\hat V$ denoting truncation of $V$. Applying this in the SVD step, a set of $k$ orthogonal components $M$ is generated via $M=P_n\hat V=U\hat\Sigma\hat V^*\hat V=U\hat\Sigma$ where $P_n$ denotes the probe matrix $P$ at the $n$-th iteration. The obtained component matrix $M$ is then used to generate the updated probes $P_{n+1}=M\hat V^*=U\hat\Sigma\hat V^*$.



\ack{We are thankful to N.\ Mahne, M.\ Zangrando and L.\ Raimondi from the PADReS group at FERMI for their contribution to the experiment leading to these results. We acknowledge the use of the IRIDIS High Performance Computing Facility, and associated support services at the University of Southampton, in the completion of this work. The research leading to these results has received funding from the European Community's Seventh Framework Programme (FP7/2007-2013) under grant agreements \mbox{n.\ 279753} and \mbox{n.\ 312284} and from Diamond Light Source Limited, the Swedish Research Council, the Knut and Alice Wallenberg Foundation, the Swedish Foundation for Strategic Research and the Swedish Foundation for International Cooperation in Research and Higher Education (STINT).}


%

\bibliographystyle{iucr}
\bibliography{ref}

\end{document}